\pdfoutput=1
\documentclass{bmcart}

\usepackage[utf8]{inputenc} 
\usepackage{xcolor}
\usepackage{lineno}
\usepackage{hyperref}

\def\includegraphics{}

\usepackage{graphicx}

\startlocaldefs
\endlocaldefs

\begin{document}

\begin{frontmatter}
	
	\begin{fmbox}
		\dochead{Methodology}
		
		
		\title{Automatic estimation of heading date of paddy rice using deep learning}
		
		
		\author[
			addressref={aff1},  
			email={cs17mtech11011@iith.ac.in}   
		]{\inits{JE}\fnm{Sai Vikas} \snm{Desai}}
		\author[
			addressref={aff1},
			email={vineethnb@iith.ac.in}
		]{\inits{JRS}\fnm{Vineeth } \snm{N Balasubramanian}}
		\author[
			addressref={aff3,aff4},
			email={fukatsu@affrc.go.jp}
		]{\inits{JRS}\fnm{Tokihiro } \snm{Fukatsu}}
		\author[
			addressref={aff2},
			email={snino@isas.a.u-tokyo.ac.jp}
		]{\inits{JRS}\fnm{Seishi } \snm{Ninomiya}}
		\author[
			addressref={aff2},
			corref={aff2},
			email={guowei@isas.a.u-tokyo.ac.jp}
		]{\inits{JRS}\fnm{Wei } \snm{Guo}}
		
		
		\address[id=aff1]{
			\orgname{Department of Computer Science and Engineering, Indian Institute of Technology - Hyderabad}, 
			\street{Kandi},                     %
			\postcode{502285}                                
			\city{Hyderabad},                              
			\cny{India}                                    
		}
		\address[id=aff2]{%
			\orgname{The University of Tokyo, Graduate School of Agricultural and Life Sciences, International Field Phenomics Research Laboratory},
			\street{Nishi-Tokyo},
			\postcode{1880002}
			\city{Tokyo},
			\cny{Japan}
		}
		\address[id=aff3]{%
			\orgname{National Agriculture and Food Research Organization, Institute of Agricultural Machinery},
			\street{1-31-1 Kannondai, Tsukuba},
			\postcode{3050856},
			\city{Ibaraki},
			\cny{Japan}
		}
		
		\address[id=aff4]{%
			\orgname{University of Tsukuba, Graduate School of Life and Environmental Sciences},
			\street{1-1-1 Ten-noudai, Tsukuba},
			\postcode{3058572},
			\city{Ibaraki},
			\cny{Japan}
		}
		
		
		\begin{artnotes}
		\end{artnotes}
		
	\end{fmbox}
	
	
	\begin{abstractbox}
		
		\begin{abstract} 
			\parttitle{Background} 
			Accurate estimation of heading date of paddy rice greatly helps the breeders to understand the adaptability of different crop varieties in a given location. The heading date also plays a vital role in determining grain yield for research experiments. Visual examination of the crop is laborious and time consuming. Therefore, quick and precise estimation of heading date of paddy rice is highly essential. 
			
			\parttitle{Results} 
			In this work, we propose a simple pipeline to detect regions containing flowering panicles from ground level RGB images of paddy rice. Given a fixed region size for an image, the number of regions containing flowering panicles is directly proportional to the number of flowering panicles present. Consequently, we use the flowering panicle region counts to estimate the heading date of the crop. The method is based on image classification using Convolutional Neural Networks (CNNs). We evaluated the performance of our algorithm on five time series image sequences of three different varieties of rice crops. When compared to the previous work on this dataset, the accuracy and general versatility of the method has been improved and heading date has been estimated with a mean absolute error of less than $1$ day.

			\parttitle{Conclusion} 
			An efficient heading date estimation method has been described for rice crops using time series RGB images of crop under natural field conditions. This study demonstrated that our method can reliably be used as a replacement of manual observation to detect the heading date of rice crops.
		\end{abstract}
		
		
		
		
	\end{abstractbox}
	%
	
\end{frontmatter}



\section*{Background}
It is an established fact that rice is one of the most important crops in the world. It feeds more than half of the world's population. Thus, a good understanding of the growth stages in rice crops would enable one to use the right amount of water, fertilizers and pesticides to ensure maximum yield. This has great economical consequences since timely and high yield of rice can potentially address the food shortage problem prevailing in many parts of the world. 

When rice paddies grow from their seeds to mature plants, they go through a variety of transformations. They develop tillers, begin to grow leaves and gradually increase in height. Then their leaf stems start bulging, which conceals the developing panicle. The panicle then starts to grow and fully emerges outside. Flowering is characterized by the exsertion of the first rice panicle in the crop \cite{Yoshida1981}. heading date is characterized together by the vegetative growth phase i.e., the time period from germination to panicle initiation and the reproductive phase, meaning the time period from panicle initiation to heading \cite{Matsuo1993}. heading date is primarily used to measure the response of the rice plant to various environmental and genetic conditions. This makes it an indispensable parameter useful to breeders and researchers. By estimating the heading date and thereby observing the heading stage, a farmer can make informed crop management decisions such as: (1) deciding the optimum amount of fertilizers and pesticides for application in the field and (2) deciding the variety of crop to be grown in the field in subsequent seasons. Meanwhile, researchers can effectively leverage the knowledge of heading stage in their experiments to understand the response of the rice plant to various genetic and environmental alterations so that they can pick the best crop variety for a particular set of environmental conditions. For instance, growth stage information has been used to determine the genetic locus which affects the regional adaptation in rice \cite{Gao2014}. Genetic modifications have been proposed to artificially control the heading date in rice crops \cite{artifical_var}. Flowering time has been controlled experimentally to enable production of crops suitable for different climates \cite{Okada2017,Yano2001}. The effect of gene interactions on traits like flowering time and panicle number has been studied \cite{gene_inter}.

For the past decade, computer vision and machine learning together have witnessed a spike in multiple research domains producing state-of-the-art results in various tasks which were previously assumed to be difficult for computers to solve. Tasks such as image classification, scene understanding and image captioning have been addressed using deep neural networks with exceptional results \cite{DBLP:journals/corr/Schmidhuber14}. Deep learning is an area of machine learning which uses high-capacity function approximators (neural networks) to recognize patterns in high dimensional data such as images. Deep learning has been successfully applied in the area of plant phenotyping in extracting traits such as plant density \cite{8309180} and plant stress \cite{Ghosal201716999}. It has also been applied in species classification \cite{moss} and detecting objects of interest such as fruits \cite{s16081222}, flowering panicles \cite{Xiong2017PanicleSEGAR}, rice spikes \cite{rspike} and wheat spikes \cite{wspike,iccvw}. For a detailed treatment of the uses of deep learning in agriculture, we encourage the readers to refer to the survey by Kamilaris and Prenafeta-Boldú \cite{KAMILARIS201870}.

Related to our task, Zhu \textit{et al.} \cite{ZHU201628} have proposed a method to observe heading stage of wheat using a two-step coarse to fine wheat ear detection method based on support vector machines (SVM).  Hasan et al. \cite{wspike} more recently used an R-CNN based object detection network to accurately detect and count wheat spikes from high definition crop images; however, this approach typically requires large image datasets with object level annotations, which is very laborious. Xiong \textit{et al.} \cite{Xiong2017PanicleSEGAR} proposed Panicle-SEG, which uses a combination of CNN and entropy rate superpixel (ERS) optimization to segment rice panicles from crop images. Since our task requires us to get an estimate of the number of flowering panicles, pixel-wise segmentation of crop images such as in \cite{Xiong2017PanicleSEGAR} is not necessary. In the context of sliding window methods, Bai \textit{et al.} \cite{rspike} used a three-stage cascade method to detect rice spikes in crop images and thereby observe the heading stage. For each patch extracted from the sliding window method, an SVM classifier is applied pixelwise to detect if the patch is a spike patch. Later, a gradient histogram method and a CNN are used to refine the classification. On the other hand, our method just requires a single pass through a CNN to detect a flowering region. This saves the computation time required to train an SVM and to apply it around each pixel in a given patch. Guo \textit{et al.} \cite{Guo2015} proposed a robust approach to detect rice flowering panicles from high definition RGB images of  field  taken  under  natural  conditions. They use a sliding window method in conjunction with an SVM classifier trained on SIFT \cite{Lowe:2004:DIF:993451.996342} features. When compared to the above studies, our approach uses a much simpler algorithm to detect flowering regions in images. Instead of using multi-step classification methods, a sliding window based mechanism is used in conjunction with a CNN to detect flowering regions in a high definition image. The number of flowering regions in an image gives a statistical estimate of the number of flowering panicles exserted. The heading date is determined by observing the date at which 50\% of the flowering panicles have been exserted. One important advantage of using a CNN is that, instead of using hand-crafted image features like SIFT, the features are automatically learnt from the data. In order to demonstrate the reliability and trustworthiness of the proposed system, GradCAM \cite{DBLP:journals/corr/SelvarajuDVCPB16}, an existing method in the literature is used to provide visual explanations for the decisions made by the CNN model used in the panicle detection algorithm. For a real-world deployable intelligent system, we believe that the explainability and transparency of the system is vital.

Our aim is to estimate the heading date in a rice crop using a fast automatic system based on computer vision and deep learning. This should eliminate the need for manual visual inspection of crops which is both tedious and time-consuming. The contributions of our work are: (1) using a deep neural network to detect flowering regions from ground-level images of paddy rice, and (2) counting the detected flowering regions to estimate the heading date of the crop. An overview of the proposed method can be seen in Figure \ref{fig1}. We evaluate the performance of our method on our dataset of five time-series RGB image sequences of three different crop varieties of rice namely, \textit{Kinmaze}, \textit{Kamenoo} and \textit{Koshihikari}. We compare our method with the manual approach to heading date measurement and observe that our automatic method estimates the heading stage with an mean absolute error of $1$ day. From the results, it can be concluded that our method has the potential to be used for estimating the yield of the crop as well as an aid in making informative crop management decisions.

\section*{Methods}

An overview of the proposed method can be seen in Figure \ref{fig1}. The input to our system is a time-series sequence of images (across different days and times) of a given crop variety taken at a particular location. For each image in the sequence, we use a sliding window mechanism to detect flowering regions. At each position of the sliding window, a CNN classifier predicts if the current window consists of a flowering panicles. In this way, we detect and count the number of flowering regions in each image. For a sequence of images taken with a single camera at a specific aspect ratio, it is easy to see that the number of flowering regions (windows) in an image is directly proportional to the number of flowering panicles present in the image. Therefore, we use the number of detected flowering regions in an image as a proxy for number of flowering panicles present. We use these region counts to draw flowering graphs and observe the heading stage.

\subsection*{Image Acquisition}

The field server system was set up in our fields at the Institute for Sustainable Agro-ecosystem Services, University of Tokyo. The setup used for image acquisition is as follows. Canon EOS Kiss X5, a digital single-lens reflex (DSLR) camera was used as part of a field server system to acquire the experimental images. The captured images were then automatically uploaded to Flickr, a free cloud service via a 3G internet connection. The uploaded images were automatically obtained by an agent system \cite{fukatsu2005} and saved into a database of National Agricultural and Food Research Organization. For the acquisition of \textit{Kinmaze} and \textit{Kamenoo} datasets, the cameras were set up at a height around  $1.5m$ from the ground. The field of view of the cameras was approximately $138 cm \times 96 cm$ (focus length 24mm) corresponding to an image resolution of $5184 \times 3456$ pixels. Using this setup, time-series images were acquired every 5 minutes from 08:00 to 16:00 between and including days 84 and 91.

For the three \textit{Koshihikari} datasets, the field of view of the cameras was approximately $180 cm \times 120 cm$ (focus length 18mm). Using this setup, the images were acquired between and including days 66 and 74. The captured images have a resolution of $5184 \times 3456$ pixels. Table \ref{tab:acqu} shows further details regarding image acquisition.

\begin{table}[h!]
	\caption{\label{tab:acqu} Details of Image Acquisition}
	\begin{tabular}{|c|c|c|c|c|}
		\hline
		              &                        & Days from     & Planting         & Number of       \\
		Dataset       & Field of view          & Transplanting & Density          & Images acquired \\ \hline
		Kinmaze       & 138 cm $\times$ 96 cm  & 84 to 91      & 28 plants $/m^2$ & 645             \\
		Kamenoo       & 138 cm $\times$ 96 cm  & 84 to 91      & 28 plants $/m^2$ & 768             \\
		Koshihikari-1 & 180 cm $\times$ 120 cm & 66 to 74      & 16 plants $/m^2$ & 680             \\
		Koshihikari-2 & 180 cm $\times$ 120 cm & 66 to 74      & 12 plants $/m^2$ & 654             \\
		Koshihikari-3 & 180 cm $\times$ 120 cm & 66 to 74      & 28 plants $/m^2$ & 721             \\ \hline
	\end{tabular}
\end{table}

\subsection*{Training Dataset}

The CNN model needs to differentiate between a flowering and a non-flowering patch. To gather the training data required to train our CNN model, we chose to annotate 500 images from the $Koshihikari-3$ dataset. Specifically, we manually drew tight bounding boxes around the flowering regions in those images. From the annotated images, we extracted 3000 patches which correspond to the annotated flowering regions. These patches are labeled with class \textit{flower} which is a positive class. Similarly, we extracted background patches randomly from the non-annotated parts of the said 500 images to obtain 3000 patches which are labeled with class \textit{non-flower} which we consider a negative class. In summary, we have a training dataset of 3000 images of positive class and 3000 images of negative class. Before training, we resize the patches to a fixed size of $224 \times 224$ pixels. Using these images, our CNN model is trained to classify a patch into one of the two classes. Figure \ref{fig2} shows examples of the patches present in the training dataset.

Generalization is an important characteristic of a machine learning model. Simply put, a model trained on one dataset should be able to perform well on similar datasets on which it wasn't trained. To assess the generalization capability of our model, we gathered training patches only from one dataset i.e., $Koshihikari-3$ and tested our model on all the five datasets.

\subsection*{Validation and Test Datasets}

We evaluate the: (1) detection performance and (2) accuracy in heading date estimation of our model separately. To evaluate the detection performance of our method, we create a validation set of images as follows. We choose 15 images from each of the five datasets mentioned in Table \ref{tab:acqu}. We pick three different time slots for choosing images: 8am - 9am, 11am - 12pm and 3pm - 4pm. We ensure that the timestamps of the chosen 15 images in any given dataset are equally distributed among these three time slots. We do this to test the robustness of our model in detecting images at various lighting conditions. From each of the 15 images, we randomly crop out a $1000 \times 1000$ portion of the image and draw tight bounding boxes around the flowering panicles present in the image. In summary, the validation set contains 75 annotated images of size $1000 \times 1000$. Note that the validation set is not used to evaluate the heading stage estimation performance, which requires counting the flowering regions. Thus, randomly cropping out a portion of the full image does not affect the evaluation method because the validation set is solely used to evaluate the detection performance of the model. 

To assess the heading stage estimation accuracy of our method, we apply our method to all the five sequences of images given in Table \ref{tab:acqu} and report the predicted heading date. In other words, we consider those five sequences as our test set.

\subsection*{Training a CNN End to End}

We train a Convolutional Neural Network (CNN) to learn the mapping between our the image patches and their labels in the training dataset. A CNN is a specially designed Artificial Neural Network (ANN) generally used to learn patterns and solve computer vision tasks from large amounts of image data. It allows for automatic feature extraction and pattern classification within its architecture. Basically, ANNs are function approximators which are generally used to learn the relationship between high dimensional input and output data. ANNs consist of several computational points called nodes connected together in the form of a directed acyclic graph. The nodes in the ANN are grouped into layers. Generally, the input data passes through one or more hidden layers sequentially before passing through the final layer to obtain the output. The choice of the number of nodes, type of nodes and number of layers constitute the architecture of the ANN. Stacking multiple hidden layers together to form a `deep' network is commonly done in order to get better representations of data. 

In the current study, we use the ResNet-50 \cite{DBLP:journals/corr/HeZRS15} architecture which is a CNN model having state-of-the-art results in image classification. For a 50-layer deep network, it is evident that we need massive amounts of data to train the network. But it is generally difficult and time-consuming to obtain massive annotated datasets especially in the agricultural domain. Therefore, we apply the widely used technique of transfer learning. We use a pretrained ResNet-50 model trained on the ImageNet \cite{Russakovsky2015} dataset which is the source domain. Now, we remove the last layer in ResNet-50 i.e., the 1000-way softmax layer and replace it with a single node sigmoid layer which gives the probability of the class being positive (flower). The weights in the model are now finetuned with data from our target domain i.e., the training dataset of patches. The process of feature extraction and classification is not separated in this case. The model is just trained end-to-end with our training data. The convolutional layers are responsible for generating the feature descriptors for the images. The sigmoid layer at the end takes these features as input and outputs the probability of the input image belonging to a positive class. The model is trained for 3 epochs using Stochastic Gradient Descent with a learning rate of 0.001 and momentum of 0.9.  

To test our model on the full images, we run a sliding window over each image. At every position of the sliding window, the model classifies the patch of the image beneath the sliding window into one of the two classes. If the model classifies the patch as a flower, then a bounding box is drawn over that sliding window as shown in Figure \ref{fig1}. 

\subsection*{Sliding Window Parameter Selection}

	In a sliding window mechanism, there are two important parameters to decide: (1) the dimensions of the window and (2) stride (step length) of the window. We have manually performed experiments on the validation dataset and empirically decided the sliding window dimensions and stride length for each dataset as shown in Table \ref{tab:detect}. The reason for having different parameters for different datasets is the fact that, despite having the camera at a fixed location above the ground, the plant height may vary for different crops. Due to the variation in plant height, the average size of flowering panicles as observed by the camera might not be consistent across different datasets. Therefore, we empirically choose the sliding window parameters separately for each dataset. 

\subsection*{Flowering Region Detection}

Since the images in each of the five datasets are in chronological order, the first step to determine the heading date is to detect the flowering regions in the images and get an estimate of the flowering panicle count. We use a sliding window mechanism to detect flowering regions in each image. The procedure of flowering region detection is described in Figure \ref{fig1}. At each position of the sliding window, the patch of the image under the window is extracted and passed through a Convolutional Neural Network (CNN). We define a flowering patch to be an image patch containing a flowering panicle. If the patch is classified by the CNN as a flowering patch, then a bounding box is automatically drawn on the boundaries of the sliding window. Once the model is trained, the model is evaluated on the test images. We use the previously mentioned five datasets as the test datasets. The procedure of testing the model on a dataset is as follows. For each image in the dataset, a sliding window is applied on the image. For each position of the sliding window, the CNN classifier detects if there is a flowering panicle in that patch. Using this process, we count the number of patches classified as flowering regions in each image.

\subsection*{Heading Date Estimation}

Once we have the flowering panicle counts for each image, we can estimate the day when $50\%$ flowering is reached which is a highly useful metric to determine the heading date of the crop. The heading stages are generally identified by percentages. Since heading stage is characterized by the exsertion of the rice panicle, the heading date can be marked as the date when 50\% of the panicles have exserted \cite{Yoshida1981}. For each dataset, we plot the cumulative distribution of detected flowering panicles against the time at which each image is captured. This allows us to find the day where 50\% of the flowering has taken place.

\subsection*{Design Decisions}
\subsubsection*{Feature Extraction vs Feature Learning}

The Scale Invariant Feature Transform (SIFT) algorithm, as used in \cite{Guo2015}, is a feature extraction algorithm. It tries to create a scale invariant representation of an image. As mentioned in the seminal paper \cite{Lowe:2004:DIF:993451.996342} by Lowe, the SIFT algorithm extracts image features that can be used for matching different images of an object. But the features extracted using the SIFT algorithm are human-engineered, in the sense that the algorithm looks for specific things like corners and edges in the image to decide its features. 

On the other hand, a deep CNN performs a series of non-linear transformations on each image to extract denser and more abstract features. The parameters of these non-linear transformations are learned by training the network with labeled data. This allows the CNN to learn distinctive features by looking at the data instead of applying some fixed mathematical transformations. Training a deep neural network end-to-end is more efficient because the learned features adapt to the task at hand i.e., classification in this case. Also, the feature extraction and classification steps are fused together in a single network. 

\subsubsection*{SVM vs Sigmoid Classification}

In the SIFT based method \cite{Guo2015}, an SVM classifier is used to classify the patches based on the SIFT features. The ResNet-50 network used in this work instead uses a one node sigmoid layer to perform binary classification i.e., it gives the probability of the input image belonging to the positive class. This layer can be seen as a logistic regression classifier. The SVM and logistic regression classifier are known to show similar performance in classification. The characteristic that makes them different is the objective function that is optimized. SVMs use a hinge loss function which tries to find the maximum margin separation between two classes of data. Logistic regression generally uses a cross-entropy loss as the cost function. The outputs of the logistic regression  classifier can be directly interpreted as the positive class probability.

\subsection*{Generating Visual Explanations}

After training and testing the CNN model, we generate visual explanations to observe the part of the image that the model looks at before detecting the presence of a flowering panicle in an image patch. For this, we take a random image from the \textit{Kinmaze} dataset and run our panicle detection algorithm which draw bounding boxes around flowering panicles in the image. Now, we randomly select a few bounding boxes and extract the patches of the image inside the bounding boxes. GradCAM \cite{DBLP:journals/corr/SelvarajuDVCPB16} is used to generate visual explanations for each image patch. In the GradCAM algorithm, we first pass the image through the CNN to get class probabilities. Since the model detected a flowering panicle in this patch, the probability of the `flower' class would be the highest. Now, the gradient of the `flower' class logit is taken with respect to each of the output feature maps of the final convolutional layer in the model. Then, global average pooling is used to calculate the weight of each feature map i.e., the importance of each feature map in causing the model to detect the presence of a flowering panicle. Finally, a heatmap is generated by taking a weighted combination of each feature map in the final convolutional layer and applying the ReLU activation function at the end. 


\section*{Results}
\subsection*{Flowering Region Detection}

We evaluate the flowering region detection performance of our method on the validation set described in the Methods section. Using the proposed method, we get the predicted flowering regions for each of the 75 images in the dataset. Note that, as shown in Figure \ref{fig3}, the ground truth annotations for the images are tight bounding boxes around the flowering panicles whereas the predicted bounding boxes are fixed size boxes detecting the flowering regions. Therefore, the standard detection evaluation metric of Intersection over Union (IoU) cannot be used to evaluate the performance of this model. Instead, we propose the following metric to evaluate the correctness of a predicted bounding box.

\begin{equation}
	Intersection\:Ratio = \frac{Area\: of\:Overlap}{Area\:of\:Predicted\:Box}
\end{equation}

\begin{table}[h!]
	\caption{\label{tab:detect} Comparison of detection performance of our model (CNN) with our previous model \cite{Guo2015} on the validation set.}
	\begin{tabular}{|c|c|c|c|c|c|c|c|c|c|}
		\hline
		Validation & No. of & \multicolumn{2}{c|}{Sliding Window} &  \multicolumn{2}{c|}{Precision} & \multicolumn{2}{c|}{Recall} & \multicolumn{2}{c|}{F1-Score}\\ \cline{3-10}
		Dataset       & images & Dimensions       & Stride & \cite{Guo2015} & CNN  & \cite{Guo2015} & CNN  & \cite{Guo2015} & CNN  \\ \hline
		Kinmaze       & 15     & 140 $\times$ 140 & 140    & 0.81           & 0.96 & 0.73           & 0.68 & 0.77           & 0.80 \\
		Kamenoo       & 15     & 160 $\times$ 160 & 140    & 0.67           & 0.84 & 0.61           & 0.71 & 0.64           & 0.77 \\
		Koshihikari-1 & 15     & 160 $\times$ 160 & 150    & 0.72           & 0.80 & 0.65           & 0.70 & 0.68           & 0.70 \\
		Koshihikari-2 & 15     & 160 $\times$ 160 & 150    & 0.72           & 0.84 & 0.73           & 0.75 & 0.72           & 0.79 \\
		Koshihikari-3 & 15     & 160 $\times$ 160 & 150    & 0.74           & 0.89 & 0.69           & 0.71 & 0.71           & 0.79 \\ \hline
	\end{tabular}
\end{table}

	Simply put, intersection ratio (IR) is the portion of the predicted bounding box which overlaps the ground truth bounding box. A predicted bounding box is considered positive if its $IR \geq 0.5$, else it is considered negative. Using this metric, we calculate the standard binary classification metrics such as Precision, Recall and F1-Score for each dataset. Table \ref{tab:detect} shows the detection results of our model on the validation set. It can be seen that our current method outperforms our previous method which used SIFT to extract features and an SVM to classify patches. From the results, it can be concluded that our current method generalizes well and has a good detection performance on images from all the five sequences. 

\subsection*{Heading Stage Estimation}
We assess the heading stage estimation performance of our method on the five image sequences mentioned in Table \ref{tab:acqu}. For each image sequence, we use our detection pipeline to detect and count the number of flowering regions in each image. Given a fixed window size, it is easy to see that the number of detected flowering regions are directly proportional to the number of flowering panicles present in an image. In other words, more the number of flowering panicles, more will be the number of flowering regions and vice versa. To evaluate this hypothesis, we have manually counted the number of flowering panicles present in each image in \textit{Kinmaze} and \textit{Kamenoo} sequences. Figure \ref{fig4} shows the comparison between the actual flowering panicle counts and the number of detected flowering regions. The Pearson Correlation Coefficient (PCC) between the ground truth panicle counts and the number of detected flowering regions was found to be 0.844 for \textit{Kinmaze} and 0.711 for \textit{Kamenoo}. These results support our hypothesis that the number of detected flowering regions are indeed a good estimate of the number of flowering panicles present. To further strengthen this hypothesis, we have plotted in Figure \ref{fig5} the change in number of flowering regions detected and the change in number of flowering panicles present. It can be seen that, in general, if the number of flowering panicles decreases at a given point, the number of detected flowering regions also decreases. Examples of images in \textit{Kinmaze} and \textit{Kamenoo} datasets and their flowering region detection outputs are shown in Figure \ref{fig6}. A similar set of images for the three \textit{Koshihikari} datasets can be found in Figure \ref{fig_add1}. To evaluate this method of estimating the heading stage, we need to manually find the heading date of the crop by visual inspection. Since the recording of the date of 50\% flowering stage is subjective and  strongly depends on the experience and intuition of the observers, we also add the dates of 1st panicle appearance in the corresponding crop as reference since normally more than 70\% of the ears will come out within the first three days after the 1st panicle appearance has been observed \cite{Matsuo1993}. Note that for paddy rice, flowering begins with panicle exsertion \cite{Yoshida1981}. Figure \ref{fig7} shows the flowering plots of \textit{Kinmaze} and \textit{Kamenoo} datasets. Figure \ref{fig_add2} shows the flowering plots for the \textit{Koshihikari-1}, \textit{Koshihikari-2} and \textit{Koshihikari-3} datasets. Table \ref{tab:redays} shows the comparison of 50\% flowering stage between field check and our proposed method.

\begin{table}[h!]
	\caption{ Comparison of proposed method and manual observation for estimation of heading stage. Formula for estimation error: (estimated) - ( field observed) 50\% flowering date.}
	\label{tab:redays}
	\begin{tabular}{|c|c|c|c|c|c|}
		\hline
		        
		              & Transplanting & 1st panicle      & 50\%flowering    & 50\%flowering & Estimation \\ 
		              &               & appearance       & dates            & dates         & Error      \\ 
		Dataset       & Dates         & (field observed) & (field observed) & (estimated)   & (days)     \\ \hline
		Kinmaze       & 05/31/2013    & 08/21/2013       & 08/24/2013       & 08/26/2013    & +2         \\
		              & day 0         & day 83           & day 86           & day 88        &            \\
		Kamenoo       & 05/31/2013    & 08/20/2013       & 08/23/2013,      & 08/24/2013,   & +1         \\
		              & day 0         & day 82           & day 85           & day 86        &            \\
		Koshihikari-1 & 05/29/2014    & 08/02/2014       & 08/05/2014,      & 08/06/2014,   & +1         \\
		              & day 0         & day 67           & day 70           & day 71        &            \\
		Koshihikari-2 & 05/29/2014    & 08/03/2014       & 08/06/2014       & 08/06/2014    & 0          \\
		              & day 0         & day 68           & day 71           & day 71        &            \\        
		Koshihikari-3 & 05/29/2014    & 08/04/2014       & 08/07/2014,      & 08/07/2014,   & 0          \\ 
		              & day 0         & day 69           & day 72           & day 72        &            \\\hline
		\multicolumn{5}{|c|}{Mean absolute error (days)} & 0.8\\ \hline
	\end{tabular}
\end{table}

\section*{Discussion}

It can be concluded from the results in Table \ref{tab:redays} that our proposed method is fairly accurate in identifying the heading stage and estimating the heading date in paddy rice. With the definition of heading date \cite{Yoshida1981} that we used, it has become quite simple to evaluate the performance of the CNN model. We have proposed a simple automatic method to observe the heading stage of rice crops. Since the observation of heading date requires an estimate of the number of flowering panicles exserted, we are not interested in accurately localizing flowering panicles in the images. Accurate localization of objects is generally done using object detection networks such as Faster R-CNN \cite{faster} which requires bounding box level annotated data for training. In other words, the images in the training data need to be annotated by drawing tight bounding boxes around the objects of interest which are flowering panicles in our case. Getting a large number of bounding box level annotated images is both time-consuming and expensive when compared to labeling an image for classification. In our work, we completely avoid this expense by using a sliding window mechanism in conjunction with a CNN classifier. The boxes predicted by our method may not always tightly localize the flowering panicles but these errors can be tolerated in our application because our end goal is not to accurately localize flowering panicles, but to observe the heading stage for which an estimate of the flowering panicle count is sufficient.

GradCAM, a visual explanation method has been used to visualize what part of the image patch the CNN model "looks at" before detecting a flowering panicle in a given patch. This visualisation would enable the model to reason its detections. Ideally, the detection of a flowering panicle in a patch should be based on the presence of flower-specific parts in the patch. The GradCAM outputs in Figure \ref{fig8} support our proposition that this is indeed the case with the proposed CNN model. The red regions in the output heatmaps represent the pixels in the patch which influenced the detection the most. It can be seen that the red regions are on the part of the patch depicting the anthesis of flowering panicle, thus supporting our claim that the model has actually learnt specific features of the flowering panicle. 

The proposed method, however, has some limitations. The current method requires high-resolution static and ground-level images of rice crop to be able to efficiently detect flowering panicles and estimate the heading date. A possible next step in this research could be to study the performance of CNNs on images taken from fully automatic Unmanned Air Vehicles (UAVs). This is because image acquisition is much simpler and faster when UAVs are used. Assessing various phenotypic traits from UAV-based images would be immensely helpful to the agricultural community owing to the simplicity of deploying drones and the ability to collect and analyze data in real time.


\section*{Declarations}
\begin{backmatter}

	

	\section*{Competing interests}
	The authors declare that they have no competing interests.

	\section*{Funding}
	This study was partially funded by SICORP Program “Data Science-based Farming Support System for Sustainable Crop Production under Climatic Change” and CREST Program “Knowledge Discovery by Constructing AgriBigData”( JPMJCR1512) from Japan Science and Technology Agency.
	
	\section*{Author's contributions}
	S.V.D. analyzed the data and interpreted results with the input of W.G. and V.N.B., S.N. supervised the entire study, W.G. conceived, designed, and coordinated the field experiments, T.F. developed the field sever and image acquisition modules for the field monitoring system. S.V.D. wrote the paper with input from all authors.
	  
	\section*{Acknowledgements}
	We thank the anonymous reviewers for their valuable comments and suggestions.
	
	\bibliographystyle{bmc-mathphys} 
	\bibliography{bmc_article}      
	
	
	

	\section*{Figures}
	
	\begin{figure}[h!]
	\includegraphics[width=0.98\textwidth]{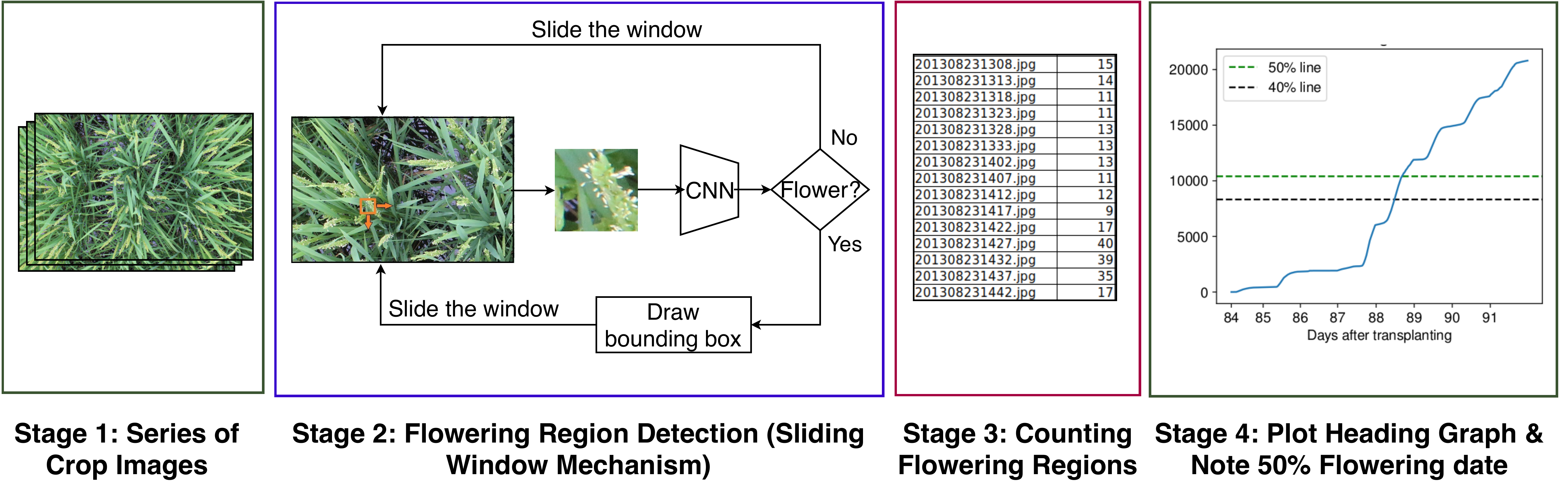}
		\caption{\csentence{Figure showing various stages of our proposed method.}
		Given a (1) time-series sequence of crop images, our sliding window + CNN method is applied on each image to perform (2) flowering region detection. Then, (3) the number of detected flowering regions are counted after which the (4) heading stage graphs are plotted.
		}
		\label{fig1}
	\end{figure}

	\begin{figure}[h!]
	\includegraphics[width=0.98\textwidth]{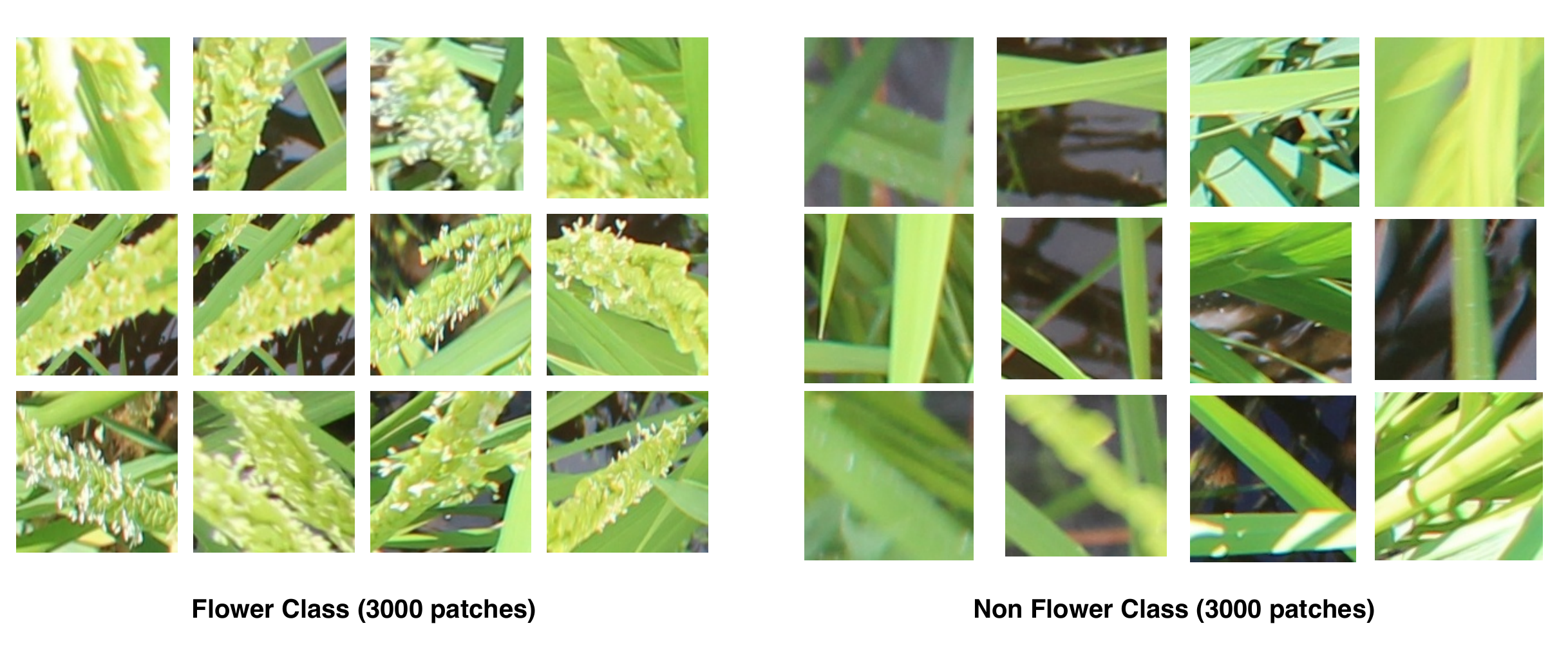}
	\vspace{-0.3in}
		\caption{\csentence{Training Data.}
			Examples of patches from the training dataset.
		}
		\label{fig2}
	\end{figure}

	\begin{figure}[h!]
	\includegraphics[width=0.95\textwidth]{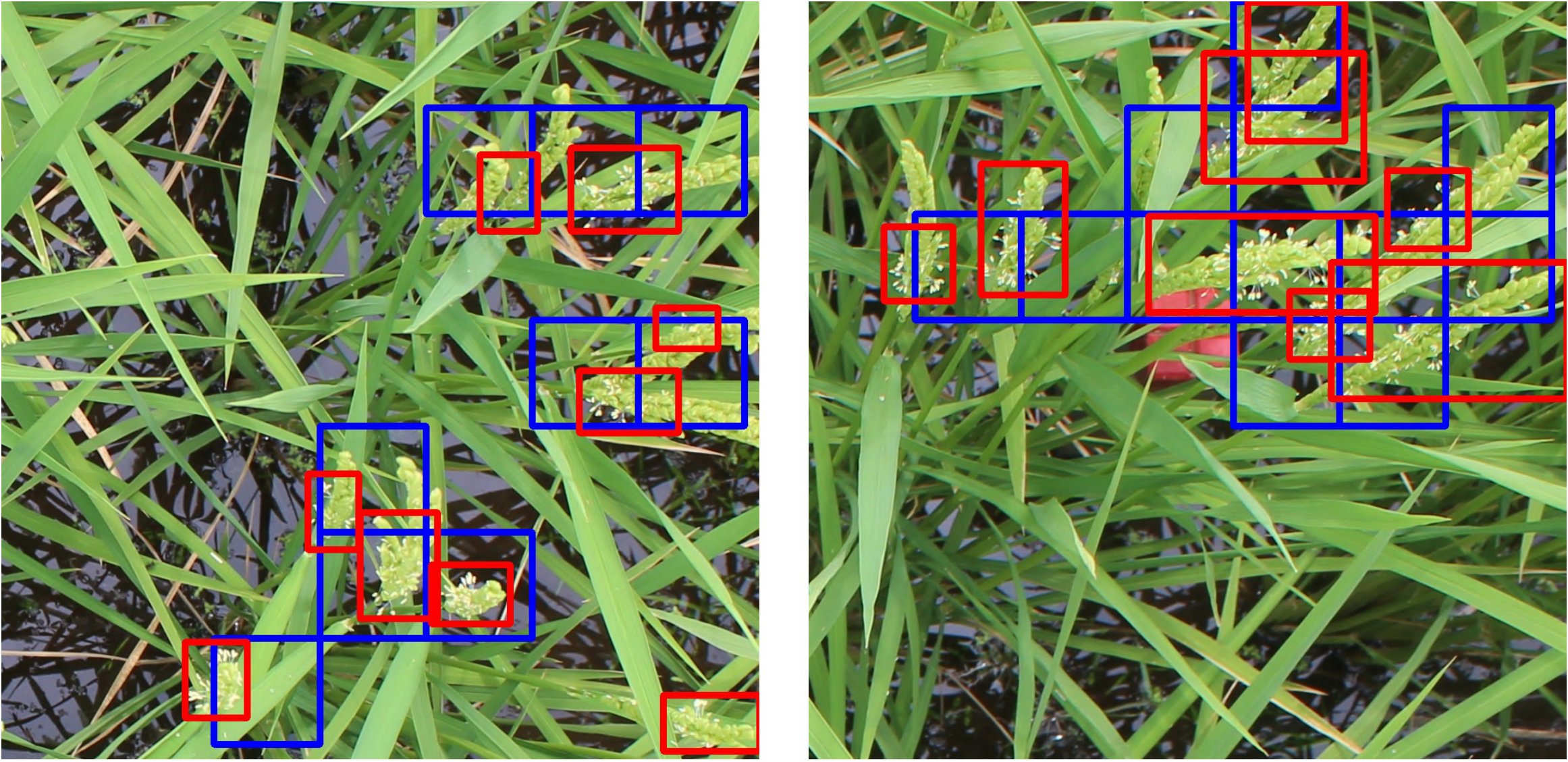}
		\caption{\csentence{Detection evaluation on validation set.}
		Examples of flowering region detection on the validation set. The ground truth boxes are shown in red and the predicted flowering regions are shown in blue.}
		\label{fig3}
		
	\end{figure}
	
	\begin{figure}[h!]
	\includegraphics[width=0.98\textwidth]{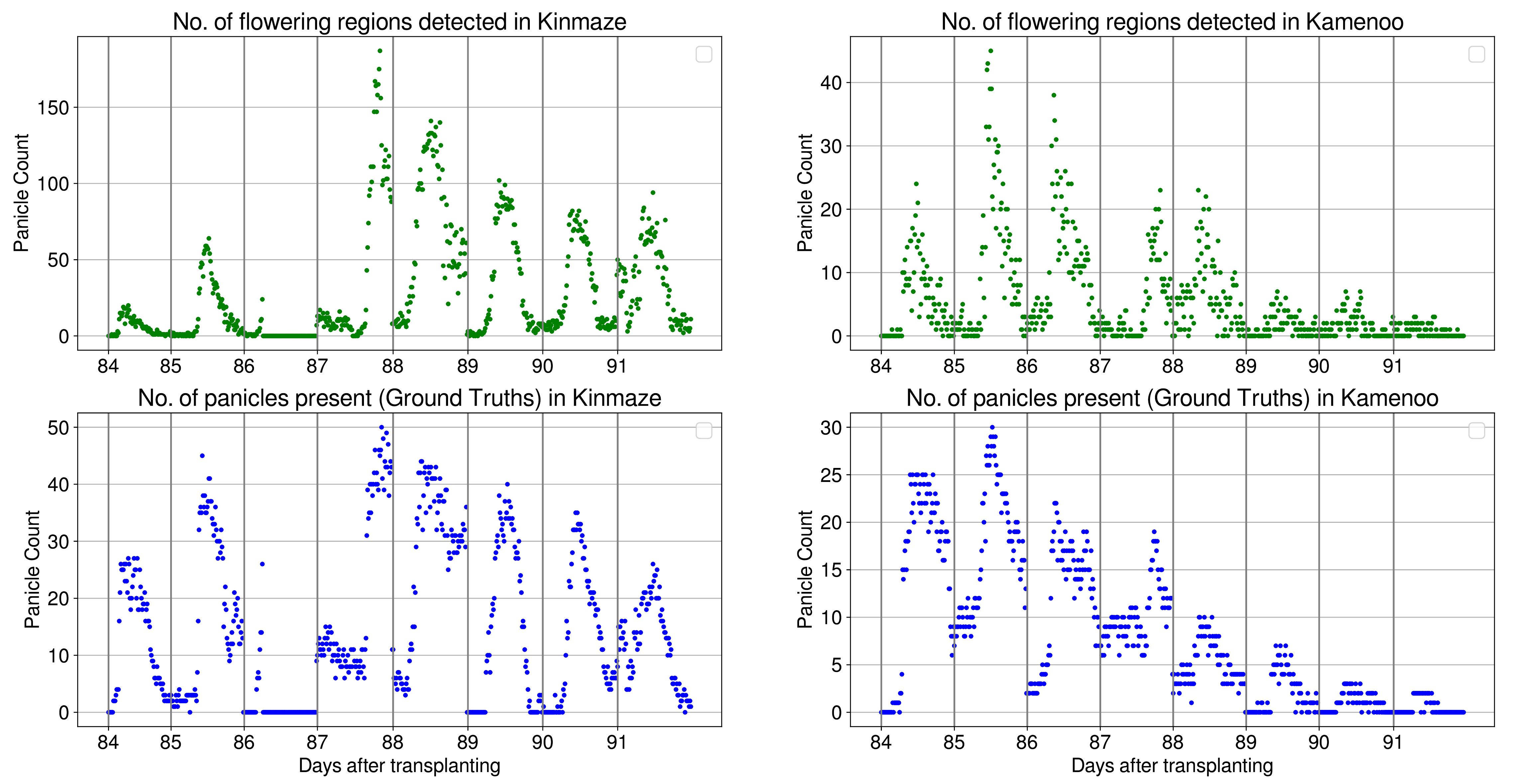}
	\caption{\csentence{Daily flowering counts. }
		Predicted flowering regions vs actual daily flowering panicle counts in Kinmaze (left) and Kamenoo (right) datasets.}
		\label{fig4}
	\end{figure}      
	   
	\begin{figure}[h!]
	\includegraphics[width=0.98\textwidth]{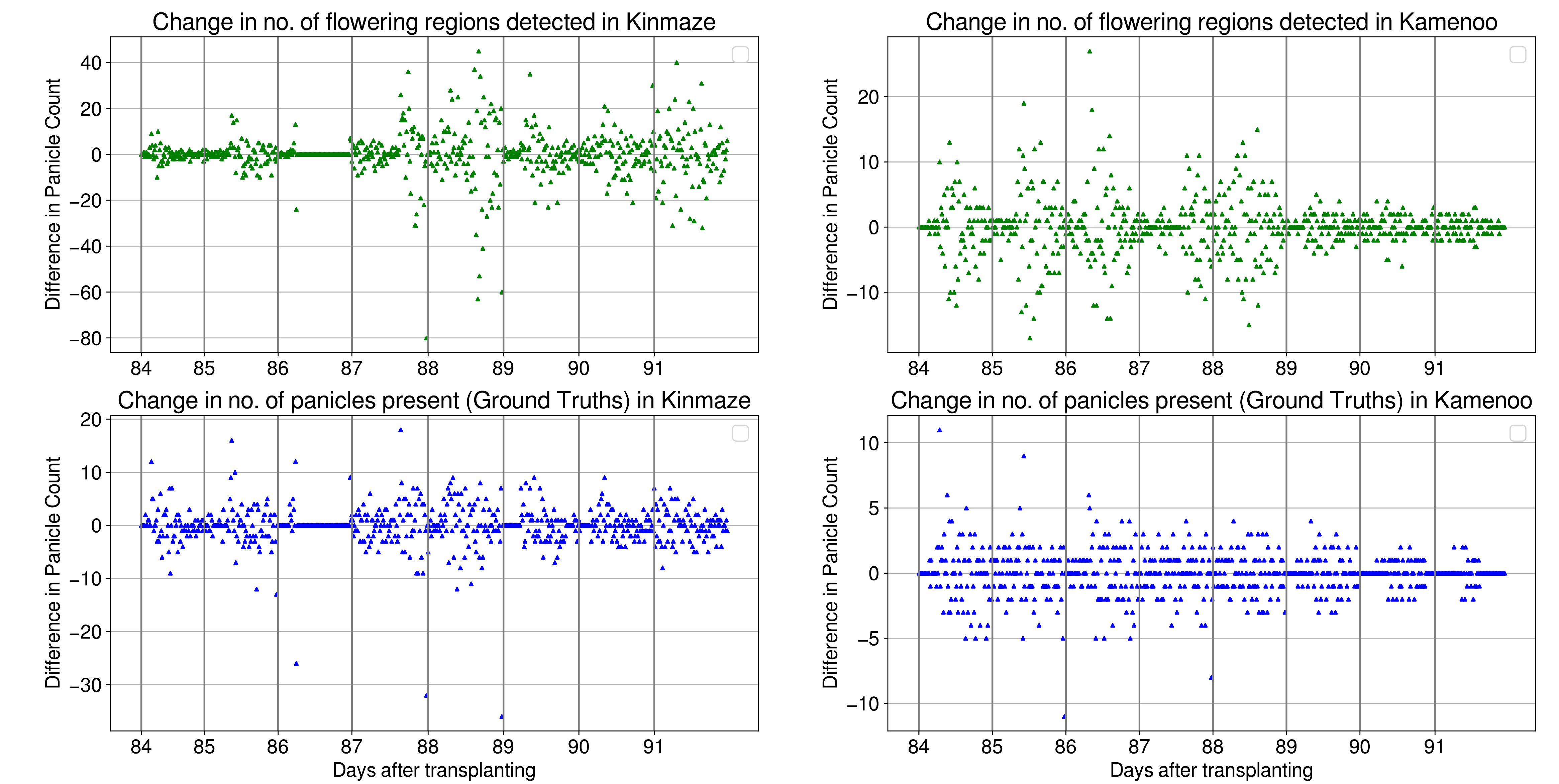}
		\caption{\csentence{Flowering Region Detection. }
		Change in number of predicted flowering regions vs change in number of actual daily flowering panicle counts in Kinmaze (left) and Kamenoo (right) datasets.}
		\label{fig5}
	\end{figure}

	\begin{figure}[h!]
	\includegraphics[width=0.95\textwidth]{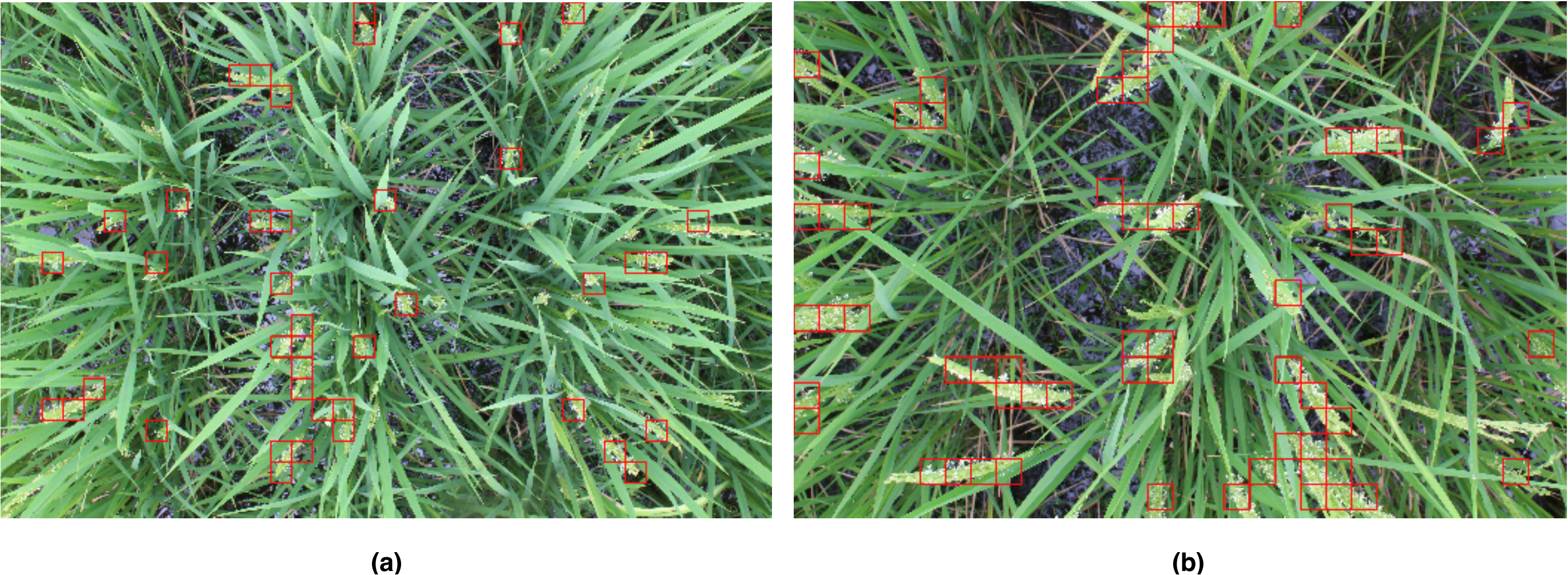}
		\caption{\csentence{Flowering Region Detection. }
		Examples of flowering region detection in datasets (a) Kinmaze (b) Kamenoo.}
		\label{fig6}
	\end{figure}

	\begin{figure}[h!]

		\includegraphics[width=0.98\textwidth]{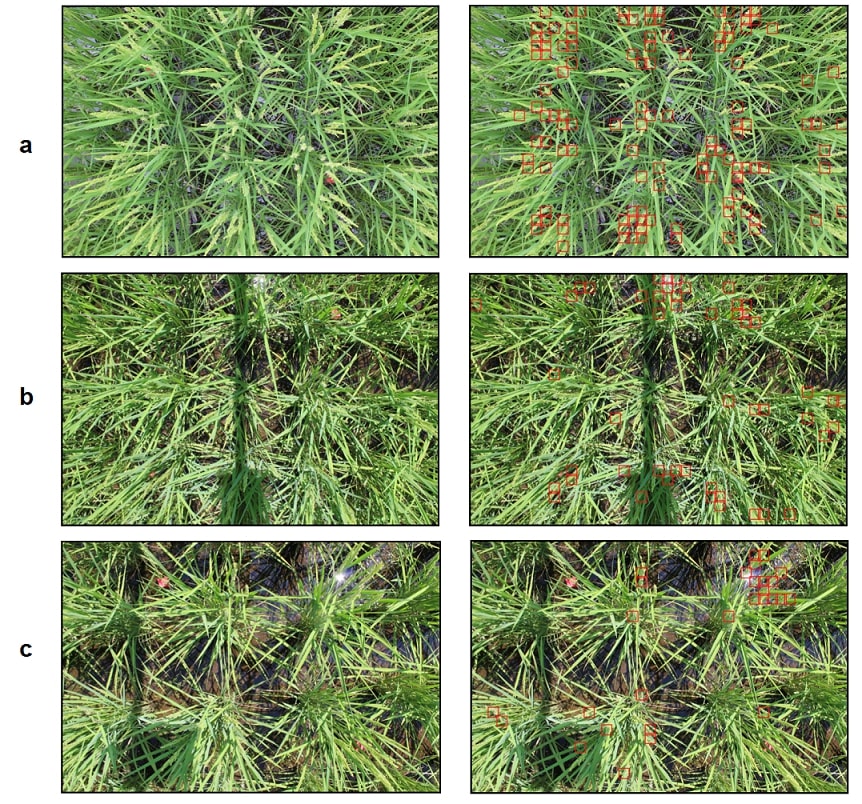}
		\caption{\csentence{Flowering Region Detection: } Examples of flowering region detection in (a) Koshihikari-1, (b) Koshihikari-2 and (c) Koshihikari-3 datasets.
		}
		\label{fig_add1}
		
	\end{figure}

	\begin{figure}[h!]
	\includegraphics[width=0.98\textwidth]{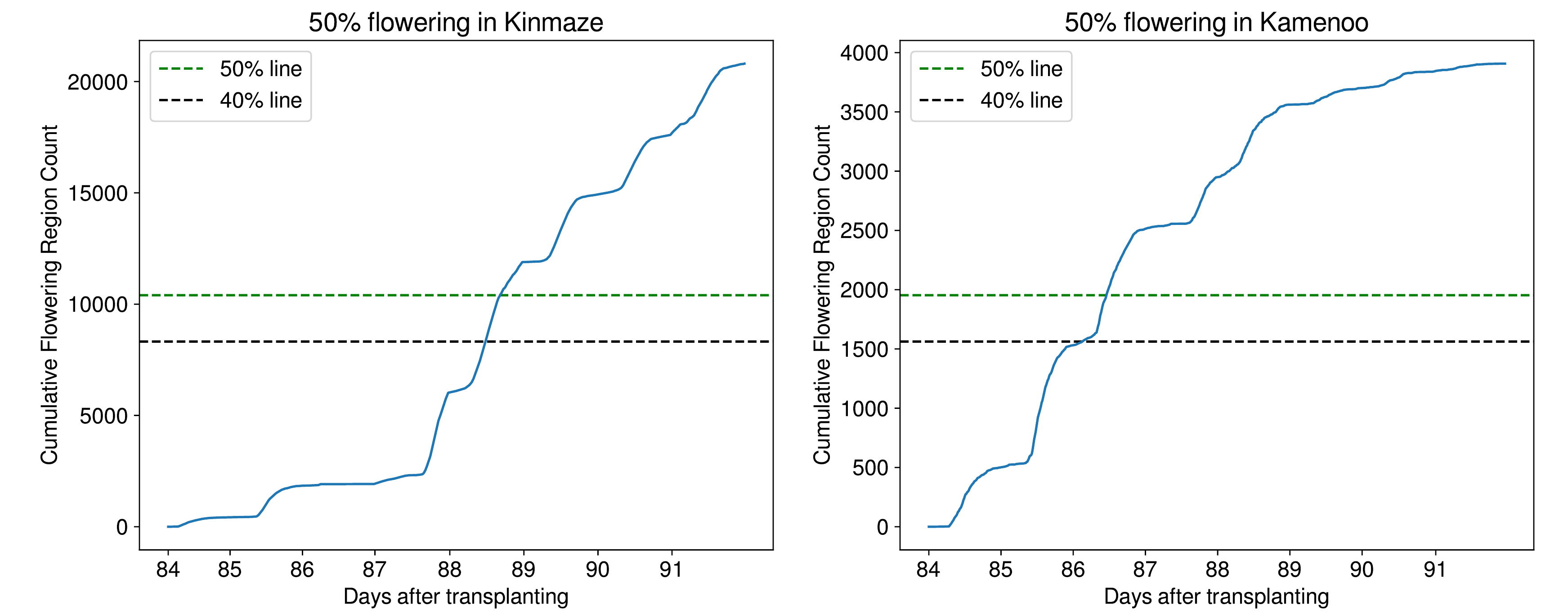}
		\caption{\csentence{Flowering stage graphs for Kinmaze and Kamenoo crops. } The estimated 50\% flowering day for Kinmaze is day 88. Similarly, the estimated 50\% flowering day for Kamenoo is day 86.}
		\label{fig7}
	\end{figure}

	\begin{figure}[h!]
		
		\vspace{-1in}
		\includegraphics[width=\textwidth]{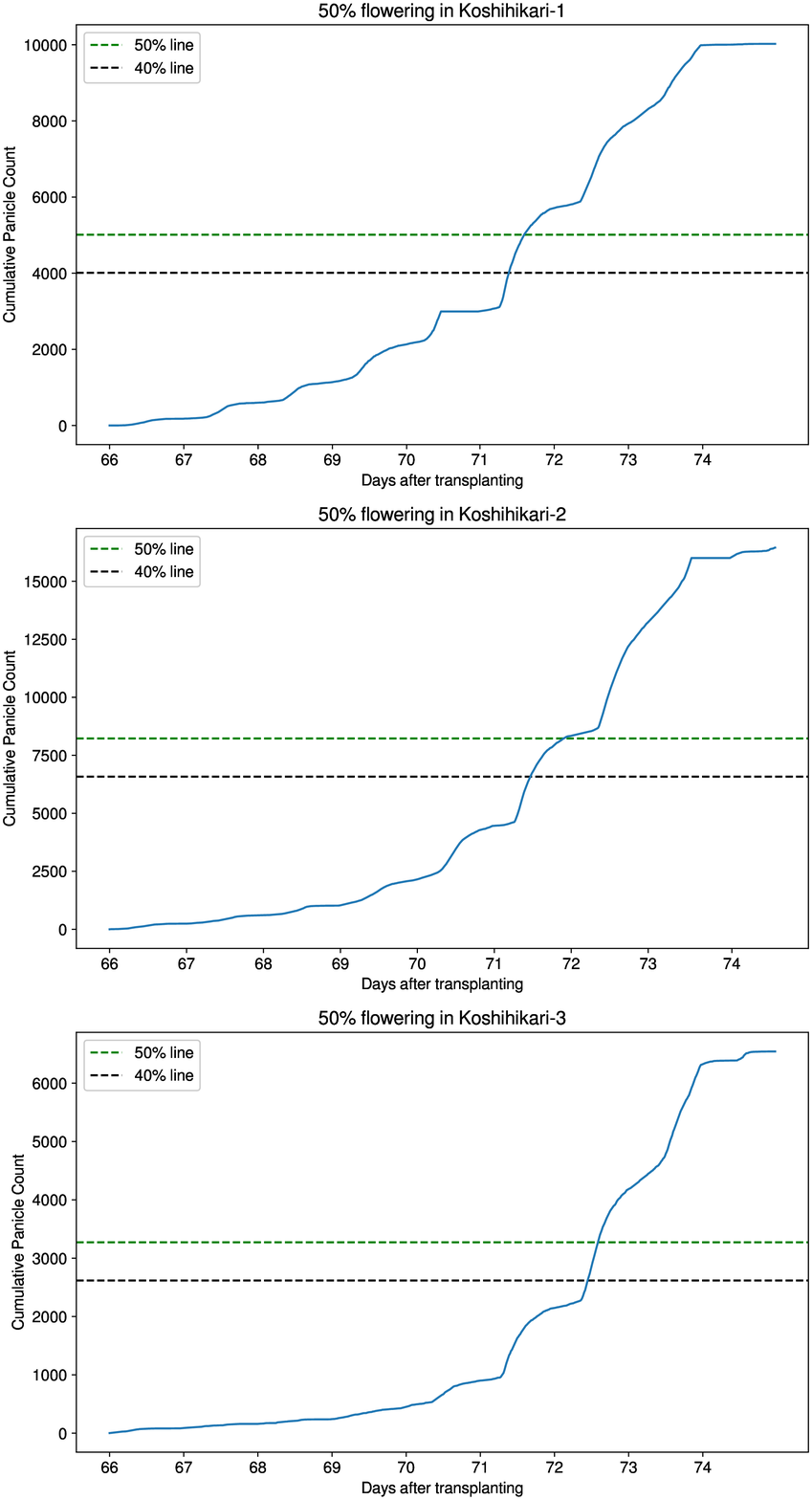}
		\vspace{-1.2in}
		\caption{\csentence{Heading stage graphs of Koshihikari-1, Koshihikari-2 and Koshihikari-3 crops.} 
		}
		\label{fig_add2}
		
	\end{figure}
	      
	\begin{figure}[h!]
		
		\vspace{-0.1in}
		\includegraphics[width=0.98\textwidth]{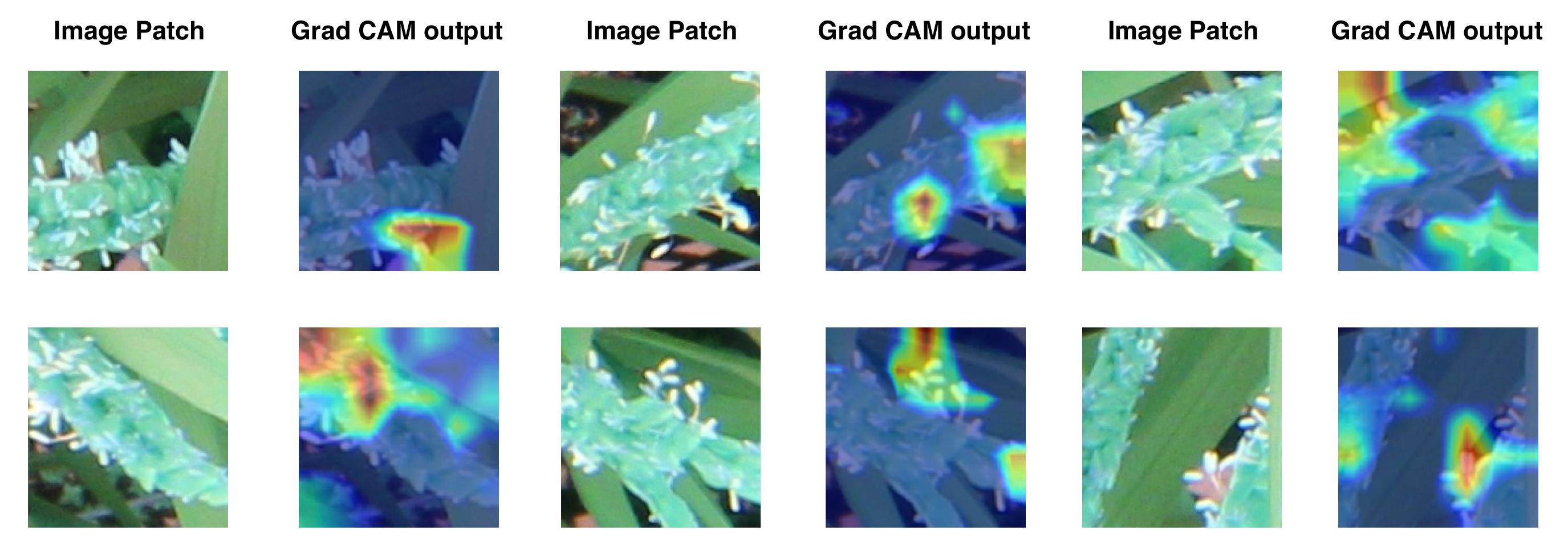}
		\caption{\csentence{Grad CAM. }
		Grad CAM outputs of flowering panicle patches with respect to the final convolutional layer of the ResNet-50 CNN are shown here. The red regions are on the part of the patch depicting the anthesis of flowering panicle, thus supporting our claim that the model has actually learnt specific features of the flowering panicle.}
		\label{fig8}
		
	\end{figure}

	

	
	

\end{backmatter}
\end{document}